\documentstyle[preprint,aps]{revtex}
\newcommand{\be}{\begin{equation}}
\newcommand{\ee}{\end{equation}}
\newcommand{\bea}{\begin{eqnarray}}
\newcommand{\eea}{\end{eqnarray}}
\newcommand{\nn}{\nonumber \\}
\newcommand{\p}{\partial}

\begin{document}
\tightenlines
\draft
\preprint{\small Preprint CGPG-97/3-6 gr-qc/9703087}

\title{Constants of motion for vacuum general relativity}

\author{Viqar Husain}

\address{Center for Gravitational Physics
and Geometry,
Department of Physics,\\
The Pennsylvania State University,\\
University Park, PA 16802-6300, USA.\footnote{
Email: husain@phys.psu.edu}}

\date{May 19, 1997}

\maketitle

\begin{abstract}

The 3+1 Hamiltonian Einstein equations, reduced by imposing two
commuting spacelike Killing vector fields, may be written as the
equations of the $SL(2,R)$ principal chiral model with certain
`source' terms.  Using this formulation, we give a procedure for
generating an infinite number of non-local constants of motion for
this sector of the Einstein equations. The constants of motion arise
as explicit functionals on the phase space of Einstein gravity, and
are labelled by sl(2,R) indices.

\end{abstract}
\bigskip
\vfill
\eject

Einstein gravity for metrics with special symmetries have been much
studied.  One such reduced system is obtained by restricting attention
to spacetimes with two commuting spacelike Killing vector fields. This
reduces Einstein gravity from a four to a two-dimensional field
theory, but still with two local degrees of freedom. This reduction
has been extensively studied starting from the initial work by Geroch
\cite{geroch}, who showed that the reduced Einstein equations have an
infinite set of hidden symmetries. Subsequent work contains other
derivations of this result, a linearized system for the evolution
equations containing a time dependent `spectral parameter', and an
identification of the algebra of the generators of the hidden symmetry
\cite{kinnchit,maison,belzak,hausern,wu}. All of this work uses the
covariant form of the reduced Einstein equations, and the
corresponding picture in the Hamiltonian formulation of Einstein
gravity has so far not been presented.

On general grounds we expect that symmetries of the covariant
equations have a phase space realization; the generators should
arise as functionals of the phase space variables that Poisson commute
with the Hamiltonian of the reduced theory - that is, as constants of
motion. In this paper we present a procedure that accomplishes this.

There are two basic approaches for obtaining a Hamiltonian formulation
of any theory obtained by symmetry reduction from a  full theory.

In one approach one starts with the covariant equations of the full
theory, reduces appropriately to obtain a set of reduced covariant
equations, and then finds a Hamiltonian formulation of the reduced
equations. {\it The Poisson bracket structure so obtained will in
general not have any relation to the fundamental Poisson bracket of
the unreduced theory}. This is so particularly for reductions to two
dimensional field theories, because many such theories are known to
have more than one Hamiltonian formulation. This feature is a known
signature of `integrability' \cite{das,fadtak}, and it is likely to
be a property of the Einstein equations reduced to two dimensions.

In the other approach one starts with the Hamiltonian formulation of
the full theory, and performs the symmetry reduction at the
Hamiltonian level. {\it This has the virtue of already having a
Poisson bracket structure, which is the natural one for the full
theory}. In the case of Einstein gravity, this would be either the
Arnowitt-Deser-Misner (ADM) Poisson bracket for the spatial metric and
its conjugate momentum (in the ADM variables), or the Ashtekar Poisson
bracket for the $SU(2)$ connection and its conjugate densitized triad
(in the Ashtekar variables) \cite{ash}.

There has been recent work on the reduced Einstein equations using
both these approaches. The first has been used in Ref. \cite{ks},
where the linearized system for these equations is used to extract
constants of motion in a Hamiltonian formulation which is based on the
one for the principal chiral model. This linearized system has also
been used to extract constants of motion in a covariant approach
\cite{nenadb}. The second has been used in Refs. \cite {lv,vh,mizo} in
the Ashtekar Hamiltonian formulation.  In Ref. \cite{mizo}, two sets
of three constants of motion are presented.  One of these sets has
also been given earlier in \cite{lv}. However the approach used in
\cite{mizo} does not allow an explicit determination on the phase
space of any more than these six conserved quantities. In
Ref. \cite{vh} the main result is a procedure for generating an
infinite number of phase space functionals that have vanishing Poisson
brackets with the reduced Hamiltonian (in a fixed time gauge). These
functionals, however, are not constants of motion because of they have
explicit time dependence.  However, because their {\it only} time
dependence is the explicit one, these functionals represent an
infinite number of solutions of the Hamiltonian evolution equations. A
question left open in Ref. \cite{vh} is how constants of motion are to
be extracted from the infinite set of solutions.  If the functionals
are simple enough, this amounts to merely identifying coefficients of
powers of time in the solutions as the constants of motion. However,
for the reduced Einstein equations it is not this simple, and an
alternative approach is needed.

In this Letter we present a simple method for obtaining an infinite
number of non-local constants of motion for the two-Killing field
reduced Einstein theory {\it directly on the gravitational phase
space}, using a form of the evolution equations derived by the author
in Ref. \cite{vh}. The method is partly based on a recursive procedure
given by Brezin et. al. \cite{bizz} for obtaining non-local conserved
charges of the principal chiral model. The charges for this latter
model were initially derived by Luscher and Pohlmeyer \cite{lp} using
a different method.

We first give an outline of the derivation of the relevant reduced
Einstein equations.  This reduction is presented in Ref. \cite{lv},
and the evolution equations in a particular time gauge (`Gowdy gauge')
are studied further in Ref. \cite{vh}, where details omitted here may
be found. The Ashtekar phase space variables are the canonically
conjugate pair $(A_a^i, \tilde{E}^{ai})$, where $i,j,\cdots=1,2,3$ are
internal $SU(2)$ indices, and $a,b,\cdots=1,2,3$ are three-space
indices. These variables satisfy the fundamental Poisson bracket
$\{A_a^i(\vec{x},t), E^{bj}(\vec{y},t) \} =i
\delta(\vec{x},\vec{y})\delta^{ij}$, and are subject to the first
class Hamiltonian, spatial-diffeomorphism and Gauss constraints
\cite{ash}. The main steps in the reduction to the two-Killing field
sector we are interested in are:

\begin{list}{}
\item{(i)} In a local cordinate system $(x,y,\theta,t)$, require that
the phase space variables depend only on one angular coordinate
$\theta$, and time $t$. This amounts to imposing two commuting
spacelike Killing vector fields $(\partial/\partial x)^a$ and
$(\partial/\partial y)^a$, and gives a 1+1 field theory on $S^1\times
R$.
\item{(ii)} Set some pairs of the phase space variables to zero:
$A_\theta^1=A_\theta^2=A_x^3=A_y^3=0$ and $\tilde{E}^{\theta
1}=\tilde{E}^{\theta 2} =\tilde{E}^{x3}=\tilde{E}^{y3}=0$. This
amounts to a partial gauge fixing, and solving the corresponding
constraints. The resulting system is first class - there is still a
Hamiltonian constraint, one remaining diffeomorphism constraint (for
the $S^1$ $\theta$ direction), and a $U(1)$ remnant of the $SU(2)$
Gauss law constraint. This is the reduction performed in \cite{lv}.
\item{(iii)} Fix a time gauge and a $\theta$ coordinate fixing
condition to obtain a reduced Hamiltonian, and then derive the
Hamiltonian evolution equations for certain combinations of the
original phase space variables \cite{vh}. The time gauge used in
Ref. \cite{vh} is the same as the Gowdy gauge \cite{gow}. The reduced
$U(1)$ Gauss law remains first class, while the spatial-diffeomorphism
constraint, together with its gauge fixing condition, form a second
class pair of constraints. Hence Dirac brackets (which in this case
happen to be the same as the usual brackets) are used to derive the
evolution equations.
\end{list}

These steps lead to the following  $1+1$ dimensional field
theory \cite{vh}. The canonical phase space variables are the pairs
$A_\alpha^I(\theta,t), E^{\beta J}(\theta,t)$, where $I,J\dots=1,2$
and $\alpha,\beta,\dots =x,y$. The Hamiltonian density is
\be
   H = {i\over \alpha}[\ \epsilon^{IJ}E^{\alpha J}\p A_\alpha^I
    + {1\over 2t} (K_\alpha^\beta K_\beta^\alpha - K^2)\ ],
\label{ham}
\ee
where $\p\equiv \p/\p\theta$,
\be
K_\alpha^\beta := A_\alpha^I\tilde{E}^{\beta I},\ \ \ K:=K_\alpha^\alpha,
\ee
and $\epsilon^{IJ}:=\epsilon^{3IJ}$ is defined from the $SU(2)$
anti-symmetric tensor $\epsilon^{ijk}$. The fundamental phase space
variables are further subject to the first class constraint $J=0$, and
the pair of second class constraints $K + \alpha=0$ ($\alpha$ is a
spacetime independent constant), and $E^{\alpha I}\p A_\alpha^I = 0$,
where
\be
J_\alpha^\beta:=\epsilon^{IJ}A_\alpha^I E^{\beta J},\ \ \
J:= J_\alpha^\alpha.
\ee

The particular  combinations of phase space variables mentioned in (iii),
which give a useful form of the evolution equations, are
\begin{equation}
L_1 = {1\over 2}(K_y^x + K_x^y), \ \ \
L_2 = {1\over 2}(K_x^x - K_y^y), \ \ \
L_3 = {1\over 2}(K_y^x - K_x^y) \ ,
\label{sl2l}
 \end{equation}
 \begin{equation}
J_1 = {1\over 2}(J_y^x + J_x^y), \ \ \
J_2 = {1\over 2}(J_x^x - J_y^y), \ \ \
J_3 = {1\over 2}(J_y^x - J_x^y) \ ,
\label{sl2j}
\end{equation}
A property of these variables is that they are invariant under the
reduced Gauss law $J=0$, obtained after the steps (ii) and (iii)
stated above. Because we will be working only with these variables in
the rest of the paper, we change notation slightly, and refer to them
as $L_i$ and $J_i$, with $i,j,\cdots=1,2,3$; from this point on the
indices $i,j,\cdots$ are {\it not} the internal $SU(2)$ indices of the
basic canonical variables.

The fundamental Poisson bracket for the canonically conjugate pair
lead to the relations
\bea
\{L_i(\theta,t),L_j(\theta',t)\} &=&
if_{ij}^{\ \ k}L_k(\theta,t) \delta(\theta,\theta'), \nn
\{L_i(\theta,t),J_j(\theta',t)\} &=&
if_{ij}^{\ \ k}J_k(\theta,t) \delta(\theta,\theta'), \nn
\{J_i(\theta,t) ,J_j(\theta',t)\} &=&
-if_{ij}^{\ \ k}L_k(\theta,t)\delta(\theta,\theta'),
\eea
where $f_{ij}^{\ \ k}$ are $sl(2,R)$ structure constants.
Therefore the $i,j,\cdots$ are naturally $sl(2,R)$ indices.

We now consider the evolution equations for the six phase space
variables $L_i$ and $J_i$. These are the only Gauss invariant degrees
of freedom that have non-trivial evolution after the (partial) gauge
fixing in (iii). Our purpose is to extract constants of motion
associated with their evolution.

The evolution equations for $L_i$ and $J_i$, derived using the
the Hamiltonian (\ref{ham}), are \cite{vh}
\be
\dot{L}_i -  J_i' = 0, \label{con}
\ee
\be
\dot{J}_i - L_i' + {2 \over \alpha t }\ f_i^{\ jk}L_j J_k = 0,
\label{cur}
\ee
where $\alpha$ is the constant (parameter) which arises from the
$\theta$ coordinate fixing condition we have used, and $\prime$ denotes
the $\theta$ derivative.

These equations may be put in a matrix form by using the
$sl(2,R)$ generators
\be
   g_1 = {1\over 2\alpha }\left( \begin{array}{cc}
         0 & 1 \\
         1 & 0 \end{array} \right)\ , \ \ \
   g_2 = {1\over 2\alpha }\left( \begin{array}{cc}
         1 & 0 \\
         0 & -1 \end{array} \right)\ ,\ \ \
   g_3 = {1\over 2\alpha }\left( \begin{array}{cc}
         0 & 1 \\
	 -1 & 0 \end{array} \right)\ ,
\label{sl2m}
\ee
which satisfy  the relations
\be
[g_i,\ g_j]= {1\over \alpha}\  f_{ij}^{\ \ k}g_k\ ,\ \ \ \ \
g_ig_j={ 1\over 2\alpha }\ f_{ij}^{\ \ k}g_k\ .
\ee
Defining the matrices
\be
 A_0 := 2 L_ig_i\ , \ \ \ \ \ \ A_1 := 2 J_ig_i\ ,
 \label{a0a1}
 \ee
the evolution equations (\ref{con}-\ref{cur}) become
\be
-\p_0 A_0 + \p_1 A_1 = 0,
 \label{cons}
\ee
\be
 \p_0 A_1 - \p_1 A_0 + {1\over t}\ [A_0, A_1] = 0,
\label{curv}
\ee
where $\p_0 \equiv \p/\p t $ and $\p_1 \equiv \p/\p \theta$.

We are now ready to derive constants of motion associated with the
evolution equations (\ref{cons}-\ref{curv}), which have been obtained,
as outlined above, using Dirac brackets after partial gauge fixing of
reduced Einstein gravity.

Consider first the simpler equations of the principal chiral model
\be
-\partial_0 A_0 + \partial_1 A_1 = 0\, ;\ \ \ \ \ \
F_{01}\equiv \partial_0 A_1 - \partial_1 A_0 + [A_0, A_1] = 0.
\label{ch}
\ee
These are equations on a flat two-dimensional Lorentzian spacetime
with topology $R^2$. There is no factor of $1/t$ multiplying the
commutator in these equations, unlike in (\ref{curv}), and therefore
the procedure we are about to review needs modification for it to be
applicable to the reduced Einstein equations. This procedure, given by
Brezin et. al.  \cite{bizz} for generating non-local conserved
currents $J_a^{(n)}$ for the first order pair of equations (\ref{ch}),
(hereafter referred to as BIZZ), uses the covariant derivative $D_a =
\p_a + A_a$, $(a,b,...=0,1)$, and proceeds as follows:

\begin{list}{}
\item (i) The first current is
\be
 J_a^{(1)}:= D_a \lambda^{(0)}=[A_a,\lambda^{(0)}],
\label{cj1}
\ee
where
$\lambda^{(0)}$ is a constant $2\times 2$ matrix. Its conservation
follows immediately from the first of equations (\ref{ch}), which in
turn implies that there is a matrix $\lambda^{(1)}$ such that $J_a^{(1)}
= \epsilon_{a}^{\ b}\p_b \lambda^{(1)}$. $\epsilon^{ab}$ is the
2-dimensional Levi-Civita tensor with
$\epsilon^{01}=+1=-\epsilon^{10}$.
\item (ii) The second current is defined by $J_a^{(2)}:=
D_a\lambda^{(1)}$.  Its conservation follows from the second of
equations (\ref{ch}) and the definition of $\lambda^{(1)}$:
\bea
\eta^{ab} \p_a J_b^{(2)} &=& \eta^{ab} D_b \p_a \lambda^{(1)} =
\eta^{ab} D_b \epsilon^c_{\ a} J_c^{(1)}  =
\eta^{ab} D_b \epsilon^c_{\ a} D_c\lambda^{(0)} \nn
&=& \epsilon^{cb}D_b D_c \lambda^{(0)} = [D_1,D_0]\lambda^{(0)}
=-F_{01} \lambda^{(0)} = 0,
 \eea
where $\eta^{ab}={\rm diag}(-,+)$ is the 2-dimensional Minkowski
metric.
\item (iii) The $n$th. current is defined by $J_a^{(n)} := D_a
\lambda^{(n-1)}$, and its conservation follows by assuming that the
$(n-1)$th. current is conserved, using steps like those in (ii). This
completes the proof by induction.
\end{list}

We now turn to the reduced Einstein equations (\ref{cons}-\ref{curv}),
and show how to obtain their conserved currents, and from these the
constants of motion.  There are two important differences from the
principal chiral model equations discussed above. The first is that
the reduced Einstein equations are on a spacetime with topology
$R\times T^3$. The second is that the curvature of $A_a$ is not zero;
there is a `source' term. Indeed, Eqn. (\ref{curv}) may be rewritten
in the equivalent forms
\be
F_{01} = (1-{1\over t})[A_0, A_1] = (1-t)(\p_0A_1 - \p_1A_0).
\label{nzero}
\ee

Our approach is based on the first two steps of the BIZZ procedure.
The first conserved current is equivalent to the equation of motion
(\ref{cons}). We  use a modified version of the second step to
obtain a {\it non-conserved} second current. This current is then
modified by adding certain `correction' terms to make it
conserved. This leads, as will see, to two sets of conserved charges,
each labelled by sl(2,R) indices. An infinite heirarchy of charges can
then be obtained from these by calculating successive Poisson
brackets, using the fundamental Poisson bracket of the gravitational
phase space variables.  The infinite set of charges so obtained have
successively higher degrees of non-locality, and are therefore
manifestly independent of one another.

The first conserved current is of course exactly the same
as (\ref{cj1}):
\be
K_a^{(1)} \equiv D_a\lambda^{(0)},
\label{k1}
\ee
where as before $\lambda^{(0)}$ is a constant matrix.
Now since $K_a^{(1)}$ is conserved, we can write
\be
K_a^{(1)} = \epsilon_a^{\ b}\p_b \lambda^{(1)} + \p_a\phi^{(1)}
\label{k11}
\ee
for some  $\lambda^{(1)}$, with  $\phi^{(1)}$  chosen to
satisfy $\Box \phi^{(1)}=0$. This change from the BIZZ procedure on
$R^2$ is necessary in order to make $\lambda^{(1)}$ a continuous
function on the circle. The potential $\lambda^{(1)}$ is given by
\be
\lambda^{(1)}(\theta,t) = -\int_0^\theta d\theta'\ K_0^{(1)}(\theta',t)
                   + \int_0^\theta d\theta'\ \p_0\phi^{(1)}(\theta',t)
                   - \int_0^t dt'\ K_1^{(1)}(0,t')
\label{lam1}
\ee
To avoid the arbitrariness introduced by $\phi^{(1)}$, we set
\be
\phi^{(1)} = {C\over 2\pi}t
\label{phi1}
\ee
where the constant $C$ is fixed by the continuity of $\lambda^{(1)}$
to the value of the first conserved charge
\be
  C\equiv Q^{(1)} = \int_0^{2\pi} d\theta\
[A_0(\theta,t), \lambda^{(0)}].
\ee

We now proceed to find the second conserved current. Set
\be
j_a^{(2)} := D_a \lambda^{(1)},
\ee
Now $j_a^{(2)}$ is {\it not} conserved because the equation of motion
(\ref{nzero}) is not a flat connection condition. Its divergence is
\bea
\eta^{ab}\p_a j_b^{(2)} &=& \eta^{ab}D_b \p_a \lambda^{(1)}
 =\epsilon^{ab} D_a (K_b^{(1)} - \p_b\phi^{(1)})  \nn
&=& \epsilon^{ab} D_a D_b\lambda^{(0)}
-\epsilon^{ab}[A_a, \p_b\phi^{(1)}]  \nn
  &=& {1\over 2}\epsilon^{ab}[F_{ab}, \lambda^{(0)}]
- \epsilon^{ab}[A_a, \p_b\phi^{(1)}] \nn
&=& (1-t) \epsilon^{ab}[\p_aA_b,\lambda^{(0)}]
 -\epsilon^{ab}[A_a, \p_b\phi^{(1)}] \nn
&=& \epsilon^{ab}\p_a [(1-t)A_b, \lambda^{(0)}] +
\epsilon^{ab}\delta_a^0 [ A_b, \lambda^{(0)}]
 -\epsilon^{ab}[A_a, \p_b\phi^{(1)}] \nn
&=& \epsilon^{ab}\p_a [(1-t)A_b, \lambda^{(0)}] +
  \epsilon^{ab}\delta_a^0 [ A_b, \lambda^{(0)}
+ {Q^{(1)}\over 2\pi}]
\label{dj2}
\eea

Now the essential observation that gives another conserved current is
that the second term on the r.h.s. of Eqn. (\ref{dj2}) may be rewritten
as a total divergence. We already know from (\ref{k11}) that
$[A_a,\lambda^{0)}]$ may be written as a total divergence. Similarly,
because $A_a$ is conserved and $Q^{(1)}$ is a constant, we can write
\be
[A_a,{Q^{(1)}\over 2\pi}]= \epsilon_a^{\ b}\p_b \bar{\lambda}^{(1)}
+ \p_a\bar{\phi}^{(1)},
\ee
where $\bar{\lambda}^{(1)}$ is defined exactly as $\lambda^{(1)}$ is in
Eqn. (\ref{lam1}), but with $[A_a,Q^{(1)}/ 2\pi]$ replacing
$K_a^{(1)}=[A_a,\lambda^{(0)}]$, and the constant
\be
\bar{C} = \int_0^{2\pi} d\theta\ [A_0(\theta,t), {C\over 2\pi}]
\ee
 replacing $C$ in (\ref{phi1}). Therefore the second term on the
r.h.s of Eqn. (\ref{dj2}) becomes
\be
  \epsilon^{ab}\delta_a^0 [ A_b, \lambda^{(0)}
+ {Q^{(1)}\over 2\pi}] =
-\delta^a_0\p_a(\lambda^{(1)}+ \bar{\lambda}^{(1)})
+\epsilon^{0a}\p_a(\phi^{(1)}+\bar{\phi}^{(1)}).
\ee

We may now identify the second conserved current. It is
\be
K^{(2)}_a\equiv j_a^{(2)}-\epsilon_a^{\ b}(1-t)[A_b,\lambda^{(0)}]
-\delta^0_a(\lambda^{(1)}+\bar{\lambda}^{(1)})
+ \epsilon_{0a}(\phi^{(1)}+\bar{\phi}^{(1)}).
\ee
Hence the second conserved charge is
\bea
Q^{(2)} &=& \int_0^{2\pi} d\theta\ K_0^{(2)}
= \int_0^{2\pi} d\theta\ \bigg\{ t\ [\lambda^{(0)},A_1(\theta,t)]
+ {\theta\over 2\pi}\Big( [A_0,C] - C - \bar{C} \Big) \nn
&&
+[\ \Big(\int_0^\theta d\theta'\ A_0(\theta',t)
+ \int_0^t dt'\ A_1(0,t')\ \Big),\ \lambda^{(0)}+ {C\over 2\pi}\ ]
 \nn
&&
 - [\ A_0(\theta,t),\ [\ \Big(\int_0^\theta d\theta'\
A_0(\theta',t),\lambda^{(0)} +
  \int_0^t dt'\ A_1(0,t')\ \Big),\ \lambda^{(0)}]\ ]
 \ \bigg\}.
\eea
Peeling off the matrices $g_i$ allows these charges to be rewritten
as two sets labelled by $sl(2,R)$ indices:
\be
Q^{(1)}_i = f_i^{\ jk}\int_0^{2\pi} d\theta\ L_j(\theta,t)\lambda_k
\ee
\bea
Q^{(2)}_i &=& -\pi(C_i + \bar{C}_i) + f_i^{\ jk}
\int_0^{2\pi} d\theta\ \bigg\{\ t\ \lambda_j J_k(\theta,t)
+ {\theta\over 2\pi}L_j(\theta,t)C_k \nn
&&
+ \Big(\lambda_k +{C_k\over 2\pi}\Big)
\Big( \int_0^\theta d\theta'\ L_j(\theta',t)
+ \int_0^t dt'\ J_j(0,t')\ \Big)
\nn
&& - L_j(\theta,t)f_k^{\ lm}\lambda_m\Big( \int_0^\theta d\theta'\
L_l(\theta',t)\ +\int_0^t dt'\ J_l(0,t')\ \Big)\
\bigg\},
\eea
where we have written the constant matrix as $\lambda^{(0)}=
\lambda_ig_i$, with constant $\lambda^i$.

The $Q^{(1)}_i$ are known from earlier work \cite{lv}. Their Poisson
brackets form the $sl(2,R)$ algebra, and so they cannot be used to
obtain new conserved charges via Poisson brackets. Indeed, they just
rotate the $Q^{(2)}_i$ among themselves.  New charges can however be
obtained by taking Poisson brackets of the $Q^{(2)}_i$ with
themselves. Since the charges are evaluated on a fixed time surface,
the usual rule
\be
\{ Q^{(2)}_i,  Q^{(2)}_j \}_t = \int_0^{2\pi}d\theta\ \bigg(
{\delta  Q^{(2)}_i\over \delta A_a^I(\theta,t)}
{\delta  Q^{(2)}_j\over \delta E^a_I(\theta,t)}
- (i\leftrightarrow j)\bigg)
\ee
for taking Poisson brackets may be used. It is easy to see that
$f_i^{\ jk} \{ Q^{(2)}_j, Q^{(2)}_k \}$ will give constants of motion,
$Q^{(3)}_i$, containing terms of one higher degree of non-locality
than $Q^{(2)}_i$; that is, they will contain terms with three spatial
integrals, two of which are nested within the full space integral.
Calculating successive Poisson brackets in this way will therefore
yield an infinite set of conserved charges $Q^{(n)}_i$, with the new
charges containing more nested spatial integrals than their
predecessors. The integer $n$ labelling the charges therefore gives
their `degree of non-locality'.

We note a few further points: (i) The method we have given may be used
to extend the BIZZ procedure for the chiral model on $R^2$ to the
model on $R\times S^1$. (ii) It does not seem easy to give a recursive
BIZZ procedure for the reduced Einstein equations because it becomes
increasingly difficult to rewrite the right hand sides of the
successive non-conserved currents as a sum of $t$ and $\theta$
derivatives.  Fortunately, as we have seen, this difficulty is
overcome by noting that the second set of conserved charges are all
that are needed to generate an infinite set of charges via Poisson
brackets.  Although it is not necessary, the same may be possible for
the principal chiral model in a Hamiltonian formulation.  (iii) The
method seems extendible to a larger class of `generalized' chiral
models $-$ those with `source' terms, where the r.h.s. of
(\ref{nzero}) can be other functions of $t, \theta, A_0, A_1$, and not
just the specific one that arises for the reduced Einstein
equation. This will of course depend on the specific form of the
source, and will not work for all sources. (iv) All the constants of
motion we have obtained, except for $Q^{(1)}$, have explicit time
dependence. This is not unexpected because the reduced Hamiltonian is
itself time dependent.  $Q^{(1)}$ is time independent because it
Poisson commutes with the time dependent part of the Hamiltonian. (v)
One can also ask for symmetries of the reduced Hamiltonian - that is,
phase space functionals, perhaps with explicit time dependence, that
Poisson commute with the reduced Hamiltonian. An infinite set of such
symmetries has been obtained in \cite{vh}.

In summary, we have presented a procedure which allows the extraction
of non-local constants of motion for vacuum Einstein gravity reduced
by imposing two commuting spacelike Killing vector fields. The
procedure relies on the resemblence of the {\it Hamiltonian} Einstein
equations with the {\it covariant} principal chiral model
equations. The method may be applicable for a class of `generalized'
chiral models, for which the curvature of the two dimensional gauge
field $A_a$ is not flat. Since the constants of motion carry $sl(2,R)$
indices, the Poisson algebra of the infinite set of constants of
motion will be an $sl(2,R)$ loop algebra. The procedure seems
applicable to the usual metric variables as well, and some work is in
progress \cite{ah}.  The explicit realization of constants of motion
we have presented may be of use for certain quantization schemes,
where one seeks representations of a classical algebra of observables.

\medskip

\noindent {\it Acknowledgements}: I would like to thank Abhay Ashtekar
for discussions and collaboration on the metric variable version of
the equations discussed here. I also thank Guillermo Mena-Marugan and
Charles Torre for helpful critical comments. This work was
supported by NSF grant PHY-9514240, the Eberly Research Funds of the
Pennsylvania State University, and by the Natural Science and
Engineering Research Council of Canada in the Mathematics Department
of the University of New Brunswick.

\end{document}